\documentclass[aps, prd, superscriptaddress, preprintnumbers, floatfix, longbibliography,nofootinbib, twocolumn, 10pt]{revtex4-2}

\usepackage{graphicx,amsfonts,amsmath,amssymb,amstext}
\usepackage{float,wrapfig}
\usepackage{psfrag, caption}
\usepackage{dsfont}
\usepackage{xcolor}
\usepackage{subcaption}
\usepackage{derivative}
\usepackage{amssymb}
\usepackage{cancel}
\usepackage{physics}
\usepackage{indentfirst}
\usepackage{amsmath}
\usepackage{comment}
\usepackage{graphicx}
\usepackage[colorlinks=true,urlcolor=blue,citecolor=blue,linkcolor=blue]{hyperref}
\hypersetup{breaklinks=true}
\usepackage{bm}
\usepackage{tcolorbox}

\graphicspath{{Figures/}} 

\newcommand{\sign}[1]{\,\mbox{sgn}\left({#1}\right)}

\DeclareMathOperator\artanh{artanh}

\definecolor{purple}{rgb}{0.8,0,0.6}
\definecolor{darkgreen}{rgb}{0.00,0.6,0.00}

\definecolor{Blue}{rgb}{0,0,0.85}

\extrafloats{100}

\begin{document}

\title{Surface states and finite size effects in triple-fold semimetals}
\date{October 3, 2025}

\author{A.Yu. Prykhodko}
\affiliation{Faculty of Physics, Kyiv National Taras Shevchenko University, 64/13 Volodymyrska st., 01601 Kyiv, Ukraine}

\author{E.V. Gorbar}
\email{gorbar@knu.ua}
\affiliation{Faculty of Physics, Kyiv National Taras Shevchenko University, 64/13 Volodymyrska st., 01601 Kyiv, Ukraine}
\affiliation{Bogolyubov Institute for Theoretical Physics, 14-b Metrolohichna st., 03143 Kyiv, Ukraine}

\author{P.O. Sukhachov}
\email{pavlo.sukhachov@missouri.edu}
\affiliation{Department of Physics and Astronomy, University of Missouri, Columbia, Missouri, 65211, USA}
\affiliation{MU Materials Science \& Engineering Institute,
University of Missouri, Columbia, Missouri, 65211, USA}

\begin{abstract}
Triple-fold or pseudospin-1 semimetals belong to a class of multi-fold materials in which linearly dispersive bands and flat bands intersect at the same point, forming triple-fold crossing points. We conduct an analytical investigation of topologically protected Fermi arc surface states and finite-size effects in three-dimensional (3D) triple-fold and doubly degenerate triple-fold semimetals in continuum low-energy models. Higher topological charge of the triple-fold crossing points leads to two Fermi arcs connecting the nodes. For a single triple-fold crossing point, we found that no term in the Hamiltonian with momentum-independent elements can open a gap, prompting us to consider doubly-degenerate triple-fold fermions, where the gap can be opened by mixing the degenerate copies. Thin films of triple-fold semimetals allow for mixing between the surface and bulk states in addition to the discretization of energy levels of the latter.
\end{abstract}

\maketitle

\section{Introduction}
\label{sec:intro}

The discovery of Dirac~\cite{Liu-Chen-Na3Bi:2014, Borisenko:2014, Liu-Chen-Cd3As2:2014, Neupane-Hasan-Cd3As2:2014, Xu-Hasan-Na3Bi:2015, Liang-Zhou-Na3Bi:2016} and Weyl~\cite{Weng-Dai:2015, Huang-Hasan-TaAs:2015, Lee-Hasan:2015, Sun-Yan:2015, Liu-Felser:2015, Belopolski-Hasan:2016, Arnold-Felser:2016b} semimetals, featuring band-crossing points or nodes with linear dispersion relation and quasiparticles described by the relativisticlike Dirac and Weyl equations, has opened up a new research field known as topological semimetals~\cite{Franz:book-2013, Wehling-Balatsky:rev-2014, Felser:rev-2017, Hasan-Huang:rev-2017, Burkov:rev-2018, Armitage:rev-2018, GMSS:book}. While each Dirac point is composed of two Weyl nodes of opposite topological charge in Dirac semimetals, Weyl nodes are separated in momentum space in Weyl semimetals. Being a condensed matter analog of Weyl fermions, Weyl quasiparticles allow one to realize several effects previously predicted in the high-energy physics setting. The celebrated chiral anomaly~\cite{Adler:1969, BJ:1969} is an illustrative example that is manifested in, e.g., negative magnetoresistivity~\cite{Nielsen:1983, Huang-Chen-TaAs:2015, Yang-Zu-NbAs:2015, Zhang-Hasan-TaAs:2016, Wang-Fang-TaP:2016, Wang-Xu-NbP:2016, Li-Xu-NbAs-NbP:2017}.

In addition to pseudospin-$1/2$ Weyl quasiparticles, spatial symmetry in crystals allows for the existence of other types of fermions with higher pseudospin and dispersion laws that have no analogs in particle physics~\cite{Bradlyn-Bernevig:2016}. In particular, the energy spectrum of the multi-fold semimetals contains nodes where several bands intersect at the same point. For example, triple-point or pseudospin-1 topological metals contain nodes where Dirac cones are intersected by an additional flat (weakly dispersive) band~\cite{Zhu-Soluyanov-TriplePointTopological-2016}.

Historically, the appearance of flat bands in two-dimensional (2D) systems was predicted a few decades ago in kagom\'{e} \cite{Syozi:1951}, dice or $\alpha-\mathcal{T}_3$ \cite{Sutherland:1986, Vidal-Doucot:1998}, and Lieb \cite{Lieb:1989} lattices. The macroscopic degeneracy of electronic flat-band states plays an important role in the thermodynamic, electronic, topological, and optical properties of the corresponding systems, see Refs.~\cite{Raoux-Montambaux:2013, Kovacs-Cserti:2016, Iurov-Huang:2018, Iurov-Huang-OpticalConductivityGapped-2023, Han-Lai:2022}. In three dimensions (3D), it was experimentally shown that CoSi~\cite{Takane-Sato:2019, Rao-Ding:2019, Sanchez-Hasan:2018, Cochran2023}, RhSi~\cite{Sanchez-Hasan:2018}, and AlPt~\cite{Schroter-Chen:2019} are multi-fold semimetals which realize triple-, four-, and six-fold nodes. Being compatible with Si technology, CoSi attracted significant attention as a material that could enhance the transport properties of electronic devices~\cite{Chen-Liang:2021, Lanzilloa-Chen:2022}.

Like Weyl semimetals, pseudospin-1 semimetals are topological, whose nontrivial topology is manifested in the presence of topologically protected surface states. In Weyl semimetals, such states are known as Fermi arcs~\cite{Vishwanath, Haldane}. The Fermi arcs connect the projections of the Weyl nodes of opposite topological charges onto the surface Brillouin zone. In multi-fold semimetals, there are also topological Fermi arc surface states. The Fermi arcs are long and chiral~\cite{Sanchez-Hasan:2018}, leading to enhanced optical response~\cite{Chang-Hasan:2019-FA-Multifold}.

Until now, the properties of the Fermi arc surface states in multi-fold semimetals were studied mostly via numerical methods using tight-binding models~\cite{Fulga2017, Hsu2022}. In this work, we develop an analytical low-energy continuum model for the Fermi arc surface states in slabs of triple-fold semimetals, where we focus on the interplay of flat and dispersive bands. We show that, in agreement with the topological charge of dispersive bands and the distribution of the Berry curvature, there are two Fermi arcs per triple-fold node: one connects the nodes and the other radiates away. While the latter is reminiscent of the long arc spanning the whole Brillouin zone observed in CoSi and RhSi~\cite{Takane-Sato:2019, Rao-Ding:2019, Sanchez-Hasan:2018, Cochran2023}, its behavior away from the nodes requires a model that is valid beyond the vicinity of crossing points, ideally in the whole Brillouin zone. In thin films, finite-size effects make the energy spectrum discrete and mix the top and bottom Fermi arcs. We show that, in such films, the Fermi arcs become separated from the surface projections of the bulk states. To address the finite-size effects on the bulk spectrum, we consider a simplified model of a doubly-degenerate triple-fold semimetal. For a thin film of this semimetal, both bulk and surface spectra are derived. Surface states show two branches, one of which is gapped. Using the obtained spectra, the density of states is analytically evaluated.

This paper is organized as follows. The Fermi arcs in the two-node model of pseudospin-1 semimetals are studied in Sec.~\ref{sec:Fermi-arcs}. Finite-size effects in thin films of triple-fold semimetals are considered in Section~\ref{sec:doubly}. Our results are discussed and summarized in Sec.~\ref{sec:summary}. Technical details of the Berry curvature calculations, wave function matching in triple-fold semimetals and vacuum, and the dependence of the spectrum on the mass parameter are presented in Appendices~\ref{sec:app-topology}, \ref{sec:app-A}, and \ref{sec:app-B}. Throughout this paper, we set $\hbar=1$.

\section{Fermi arcs in two-node model}
\label{sec:Fermi-arcs}

\begin{figure}[t]
    \centering
    \includegraphics[width=0.5\textwidth]{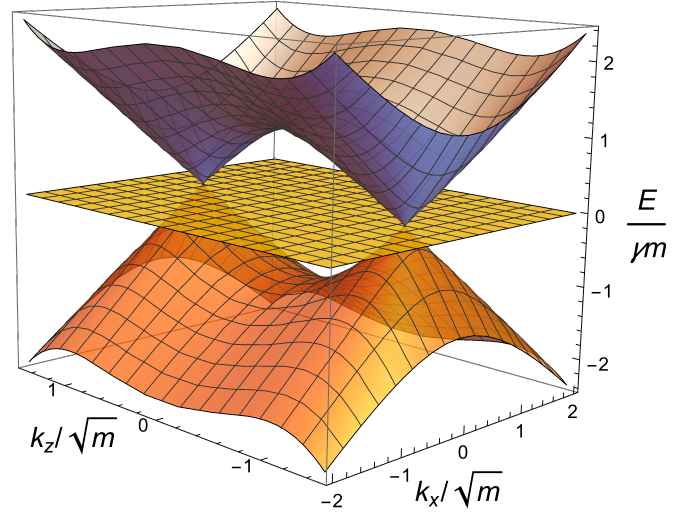}~
    \caption{
    The energy dispersion given in Eq.~\eqref{fa-bulk-spectrum} in the model of 3D pseudospin-1 semimetal with two nodes as a function of $k_x$ and $k_z$ at $k_y=0$. We set $v_F=\gamma\sqrt{m}$.
    }
    \label{fig:Okugawa}
\end{figure}

In Weyl semimetals, where the presence of an even number of Weyl nodes is a defining feature~\cite{Nielsen:1981a, Nielsen:1981b, Nielsen:1981c}, the Fermi arc surface states connect the projections of the bulk Weyl nodes of opposite chiralities (topological numbers). To have a vanishing topological charge in the whole Brillouin zone, we also consider a model with two triple-fold nodes of opposite topological charges. We use the following low-energy model describing two triple-fold nodes:
\begin{equation}
    \mathcal{H}_2=\begin{pmatrix}
        0 & i v_F k_x & -i v_F k_y\\
        -i v_F k_x & 0 & i\gamma(k_z^2-m)\\
        i v_F k_y & -i\gamma(k_z^2-m) & 0\\
    \end{pmatrix}.
\label{Hamiltonian-two-nodes}
\end{equation}
Its energy bands are
\begin{equation}
\label{fa-bulk-spectrum}
    E_0=0,\quad E_\pm=\pm\sqrt{v_F^2(k_x^2+k_y^2)+\gamma^2(k_z^2-m)^2}
\end{equation}
with the triple-fold nodes located at $k_z=\pm\sqrt{m}$. We plot this energy spectrum in Fig.~\ref{fig:Okugawa} as a function of $k_x$ and $k_z$. Note that Hamiltonian \eqref{Hamiltonian-two-nodes} defines a toy model which, while capturing the properties of triple-fold nodes, does not directly describe real materials such as CoSi. Yet, its simplicity is appealing because it allows us to get a glimpse into the main features of Fermi arcs and triple-fold fermions.

Unlike Weyl semimetals whose nodes have topological charge $\pm1$, the triple fold nodes of the model \eqref{Hamiltonian-two-nodes} have charges $\pm2$. This can be shown by calculating the flux of the Berry curvature through spheres surrounding each of the triple-fold nodes. The Berry curvature for Hamiltonian given in Eq.~\eqref{Hamiltonian-two-nodes} reads $\vec{\Omega}_{\pm} = \pm \vec{q}/\rho^3$ with $\rho=\sqrt{v_F^2(k_x^2+k_y^2)+\gamma^2(k_z^2-m)^2}$ and $\vec{q} = \left\{2\gamma v_F k_z k_x, 2\gamma v_F k_z k_y, \gamma(k_z^2 -m) \right\}$, see Appendix~\ref{sec:app-topology} for details. By calculating the flux of $\vec{\Omega}_{\pm}$ through the surfaces surrounding the nodes, we obtain the topological charge $\pm 2$.

We consider Fermi arcs at the interface between a triple-fold semimetal and vacuum. The latter is modeled as a wide-gap insulator. To open a gap, we substitute $m\rightarrow - M$ in Hamiltonian (\ref{Hamiltonian-two-nodes}). Since the flat band still remains at zero energy, we shift the energy spectrum so that the Fermi energy is in the middle of one of the two band gaps by adding $\gamma M/2$. Thus, we obtain the following Hamiltonian for an insulating phase:
\begin{equation}
\label{Twonodevac}
\mathcal{H}_2^{\text{gap}}=\mathcal{H}_2\bigg|_{m\rightarrow-M}+I_3\frac{\gamma M}2,
\end{equation}
where $I_3$ is the unit $3\times 3$ matrix. The eigenvalues of the above
Hamiltonian are
\begin{equation}
\label{eq:fa-eps-vacuum}
    E_0=\frac{\gamma M}{2},\,\,\, E_\pm=\frac{\gamma M}{2}\pm\sqrt{v_F^2(k_x^2+k_y^2)+\gamma^2(k_z^2+M)^2}.
\end{equation}

Hamiltonians (\ref{Hamiltonian-two-nodes}) and (\ref{Twonodevac}) allow us to investigate the topological surface states of pseudospin-1 semimetals and analyze their properties. For definiteness, we assume that the triple-fold semimetal is situated at $y<0$ and the vacuum is at $y>0$. The system is infinite in the $x$- and $z$-directions. In addition, we examine slabs of finite thickness $L$ that are surrounded by vacuum on both sides.

\subsection{Wave function matching}
\label{sec:Fermi-arcs-matching}

We begin with the simpler case of a semi-infinite semimetal. Given that Hamiltonian (\ref{Hamiltonian-two-nodes}) is linear in $k_y$, naively, one would expect that straightforward matching of every component of the wave function at the semimetal-vacuum boundary should immediately provide the dispersion relation for Fermi arc states. However, because of the flat band, the situation is more involved for triple-fold fermions, see, also, Ref.~\cite{Mandal2020}.

In the model with Hamiltonian \eqref{Twonodevac}, vacuum flat band states cannot be matched with the bulk solutions at the interface for $E \ne \gamma M/2$. Consequently, the general solution contains one arbitrary constant related to the solution in vacuum and two in the semimetal. Therefore, matching of three-component wave functions yields a system of three equations for three unknown coefficients. Our analysis in Appendix~\ref{sec:app-A} shows that this system has no non-trivial solutions.

The issue with matching arises due to an insufficient number of linearly independent solutions in vacuum to satisfy the matching conditions for a three-component wave function. To provide an additional solution, we consider a more realistic weakly-dispersing flat band by adding the following term to the Hamiltonian:
\begin{equation}
    \mathcal{H}^{\text{quad}}=-\delta k_y^2I_3,
\end{equation}
where $0<\delta \ll \gamma$ is a constant. [For simplicity, we ignore other quadratic terms, see Ref.~\cite{Ni-Wu-GiantTopologicalLongitudinal-2021}, for the full Hamiltonian of CoSi.] This term modifies the energy spectrum in vacuum as follows
\begin{eqnarray}
\label{eq:fa-eps-vacuum-1}
E_0 &=& \gamma\frac M2-\delta k_y^2,\\
\label{eq:fa-eps-vacuum-2}
E_\pm &=& \gamma\frac M2-\delta k_y^2\pm\sqrt{v_F^2(k_x^2+k_y^2)+\gamma^2(k_z^2+M)^2}.
\end{eqnarray}

We have the following eigenstate corresponding to $E_0$ in vacuum:
\begin{equation}\label{zerostate}
\psi_0=\frac{1}{\sqrt{v_F^2k_x^2+v_F^2k_y^2+\gamma^2(k_z^2+M)^2}}\begin{pmatrix}
        \gamma(k_z^2+M)\\
        v_Fk_y\\
        v_Fk_x\\
    \end{pmatrix},
\end{equation}
where $k_y=\pm i\sqrt{(\gamma M -2E)/(2\delta)}$ with $E<\frac{\gamma M}{2}$. In the limit $\delta\rightarrow0$, we obtain
\begin{equation}
\label{zerostate-flat}
    \psi_0=\begin{pmatrix}
        0\\
        1\\
        0\\
    \end{pmatrix}+O(\sqrt{\delta}).
\end{equation}
For $E_{\pm}$ bands, the effect of the new quadratic term is negligible. In contrast, the correction has a qualitatively different impact on the flat band $E_0$: a nonzero $\delta$ removes its degeneracy, inducing weak dispersion, such that for some values of $k_y$, the $E_0$ band states can attain the same energy as the dispersive eigenstates in the semimetal. Therefore, due to the weak dispersion of the flat band, states in the semimetal and vacuum can be matched. According to Eq.~(\ref{zerostate-flat}), the second component of the wave function accounts for flat band states, and, in the case of an ideally flat band $(\delta=0)$, matching conditions are reduced to matching of only the first and third components.

Heuristically, this result also follows from requiring vanishing current through the interface $j_y =\psi^{\dag} (\partial_{k_y}H) \psi$, which, as follows from Eq.~\eqref{Hamiltonian-two-nodes}, depends only on two components of the wave function $\psi$ corresponding to the dispersive bands. While this is straightforward to show for the form of the Hamiltonian used in Eq.~\eqref{Hamiltonian-two-nodes}, this statement also holds for an arbitrary representation of the pseudospin matrices. To show this, we use the fact that different representations are connected by a unitary transformation $U$ leading to the normal component of the current $j_y=\psi^\dagger U (\partial_{k_y}H) U^\dagger\psi$. The transformed current operator $US_{1y}U^\dagger$ has three eigenvectors corresponding to the dispersive conduction $\psi^{U}_+$ and valence $\psi^{U}_-$ bands, as well as the flat band $\psi^{U}_0$. The coordinate-independent part of the wave functions in our system can always be represented as a combination of these eigenvectors $\psi^{U} = \sum_{j=\pm,0}c_j \psi^{U}_{j}$. Since, by definition, $US_{1y}U^\dagger \psi^{U}_0 = 0$, the corresponding component of the wave function does not affect the normal component of the electric current. Therefore, in agreement with our discussion at the beginning of this section, it should not enter the boundary conditions.

\subsection{Surface states}
\label{sec:Fermi-arcs-states}

\begin{figure}[t]
\centering
\subfloat[$E=0.2\,\gamma m$ \label{arc-E=0.2}]{\includegraphics[width=0.25\textwidth]{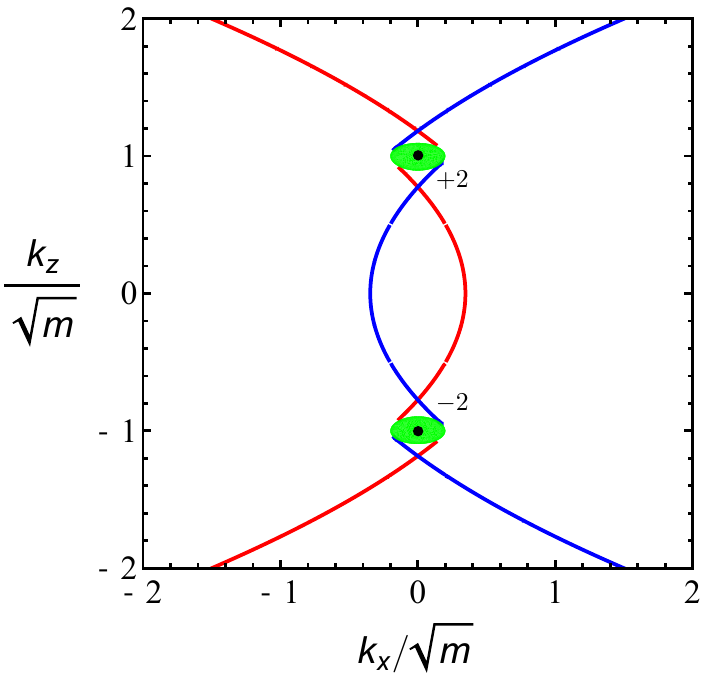}}
\subfloat[$E=0.6\,\gamma m$ \label{arc-E=0.6}] {\includegraphics[width=0.25\textwidth]{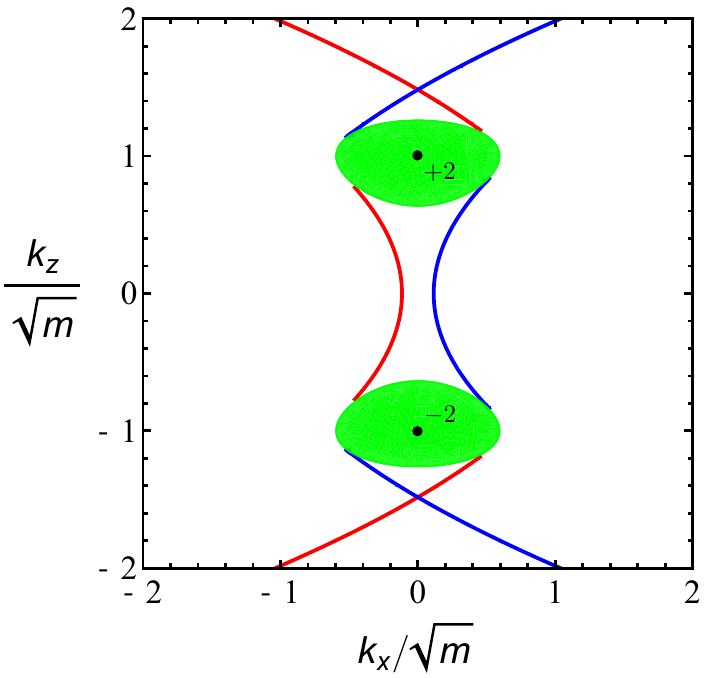}}
\\
\subfloat[$E=-0.2\,\gamma m$]{\includegraphics[width=0.25\textwidth]{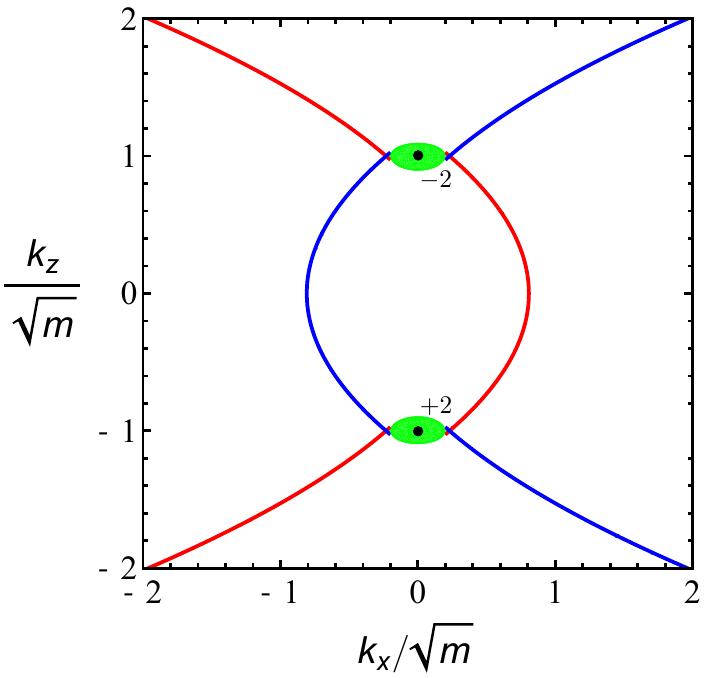}}
\subfloat[$E=-0.6\,\gamma m$]{\includegraphics[width=0.25\textwidth]{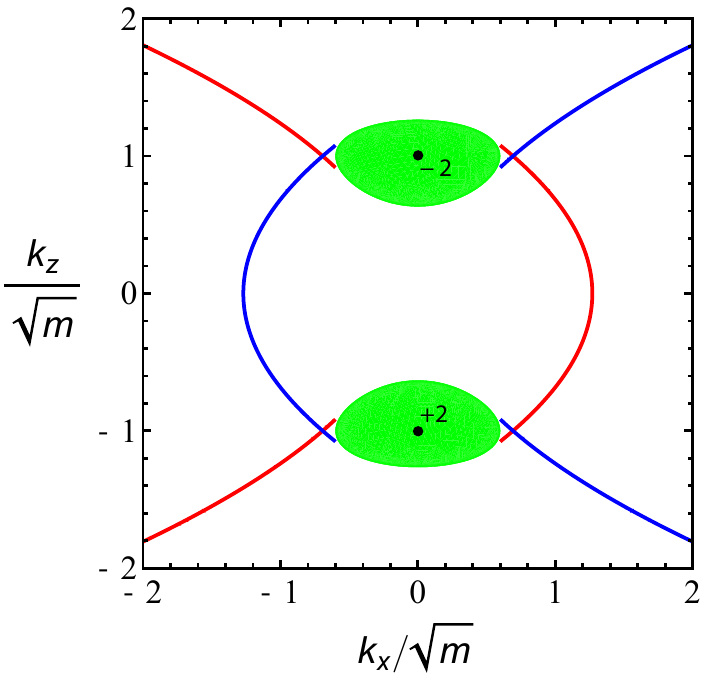}}
\caption{
Fermi arcs at the semimetal-vacuum interface for $v_F=\gamma\sqrt{m}$. Green regions represent the projections of the bulk spectrum onto the surface Brillouin zone. Blue curves correspond to the bottom surface states, while red curves correspond to the top surface. Black dots and numbers $\pm2$ mark triple-fold crossing points and the corresponding topological charges.
} \label{orc}
\end{figure}

To find the Fermi arcs, we use the approach described in Sec.~\ref{sec:Fermi-arcs-matching}. The energy dispersion of the top surface Fermi arc states is defined by the following characteristic equation:
\begin{equation}\label{Mabel}
    \frac{v_F^2k_x^2-E^2}{Eq+k_x\gamma(k_z^2-m)}=\frac{\left(E-\gamma M/2\right)^2-v_F^2k_x^2}{(E-\gamma M/2)p-k_x\gamma(k_z^2+M)},
\end{equation}
where $v_Fp = \sqrt{\gamma^2\left(k_z^2+ M\right)^2+v_F^2k_x^2-\left(E -\gamma M/2\right)^2}$ and $v_Fq = \sqrt{\gamma^2\left(k_z^2-m\right)^2+v_F^2k_x^2-E^2}$. Coefficients $p$ and $q$ determine the inverse decay lengths in vacuum and semimetal, respectively.

In the limit $M\rightarrow\infty$, Eq.~\eqref{Mabel} simplifies as
\begin{equation}\label{dipper}
    \frac{v_F^2k_x^2-E^2}{Eq+k_x\gamma(k_z^2-m)}=-\frac{v_F}{\sqrt{3}},
\end{equation}
leading to
\begin{equation}\label{true} E^2=\frac14\left[\sqrt{3}v_Fk_x+\gamma(k_z^2- m)\right]^2
\end{equation}
with the additional condition that $\sqrt{3}(v_F^2k_x^2-E^2)+v_Fk_x\gamma(k_z^2-m)$ is positive, which corresponds to surface states with positive energy on the top surface.

We present the Fermi arcs obtained from Eq.~\eqref{true} in Fig.~\ref{orc}. One can see that there are Fermi arcs connecting the projections of the bulk spectrum onto the surface~\footnote{Fermi pockets shown by green regions in Fig.~\ref{orc} transform into points, i.e., triple-fold nodes at $E\to0$.} (at $k_z^2<m$). In addition, there are Fermi arcs that go away from the projections of the Weyl nodes onto the surface Brillouin zone (at $k_z^2>m$). We speculate that these arcs span the entire Brillouin zone. However, to determine whether these Fermi arcs extend throughout the whole surface Brillouin zone, one should go beyond the effective low-energy model. We leave this question for further studies. As follows from Eq.~\eqref{dipper}, the Fermi arcs have a parabolic shape: the parts between the nodes and outside the nodes are represented by the shifted parts of the same parabola that is cut perpendicular to its symmetry axis. Fermi arcs corresponding to the top (semimetal is in the $y<0$ subspace) and bottom (semimetal is in the $y>0$ subspace) surfaces are related by the mirror $k_x \to -k_x$ symmetry. In addition, both arcs are symmetric with respect to the $k_z=0$ axis. Similar arcs were obtained numerically in Ref.~\cite{Nandy-Roy:2018}. On the other hand, Fermi arcs spanning outside the triple-fold nodes $k_z^2>m$ are absent in a symmorphic toy model considered in Ref.~\cite{Hsu2022}.

The presence of two Fermi arcs follows from the topological charge $\pm 2$ of the triple-fold crossing points. Furthermore, by calculating the Chern number $\mathcal{C}(k_z)$ for the planes with fixed $k_z$, we find $\mathcal{C}(k_z) = \sign{m-k_z^2}$. Therefore, since a nonzero Chern number is an indicator of topologically protected edge states, the model at hand can support Fermi arcs both between the crossing points and outside of them, which is indeed observed in Fig.~\ref{orc}.

The negative-energy Fermi arcs have a different shape in our model, cf. top and bottom panels in Fig.~\ref{orc}. This asymmetry originates from the particle-hole symmetry breaking of the Hamiltonian (\ref{Twonodevac}). Note that particle-hole symmetry could be preserved for a three-component Hamiltonian only if one band is completely flat and is located at zero energy. Therefore, gap opening necessarily breaks the particle-hole symmetry.

\subsection{Hybridization of surface states in thin films}
\label{sec:finite-size1}

In the previous section, we considered a semi-infinite triple-fold semimetal. Real crystals are, of course, finite and are commonly grown as films. In sufficiently thin films, the overlap between the surface states can no longer be neglected and, as we show in this section, leads to the hybridization of the Fermi arcs at opposite surfaces of the film. The hybridization of the surface states in Weyl and Dirac semimetals is discussed in Refs.~\cite{Benito-Matias-Molina-SurfaceStatesTopological-2018, Pal-Cook-FinitesizeTopologicalPhases-2025}.

Let us consider a slab of pseudospin-1 semimetal, which extends infinitely in the $x$- and $z$-directions and is contained within $-L/2 <y< L/2$ region. The electron states in the region outside the semimetal are described by the gapped Hamiltonian \eqref{Twonodevac}.

For the surface Fermi arc states, wavevector components $k_x$ and $k_z$ are good quantum numbers, and $k_y$ is purely imaginary with its sign chosen such that wave functions exponentially decrease at infinity. This gives the following characteristic equation for the surface states:
\begin{eqnarray}
    \label{film-char-eq}
    \coth(qL) &=& -\frac{q\left(E -\gamma\frac M2\right)}{2E p}-\frac{E p}{2\left(E -\gamma\frac M2\right)q} \nonumber\\
    &+& \frac{k_x^2\gamma^2(M+m)^2}{2\left(E -\gamma\frac M2\right)E pq}.
\end{eqnarray}

Let us examine the relation in \eqref{film-char-eq} in the limit of an infinitely large gap $M\rightarrow\infty$. Then, $p\rightarrow \sqrt{3}\gamma M/(2v_F)$ and we obtain
\begin{equation}\label{Farcs}
    \coth(qL)=
    \frac{2E^2-3v_F^2k_x^2+\gamma^2(k_z^2-m)^2}{2\sqrt{3}Ev_Fq}.
\end{equation}

\begin{figure*}[t]
    \subfloat[$E=0.2\, \gamma m$, $L=7/\sqrt{m}$ \label{detached}] {\includegraphics[width=0.3\textwidth]{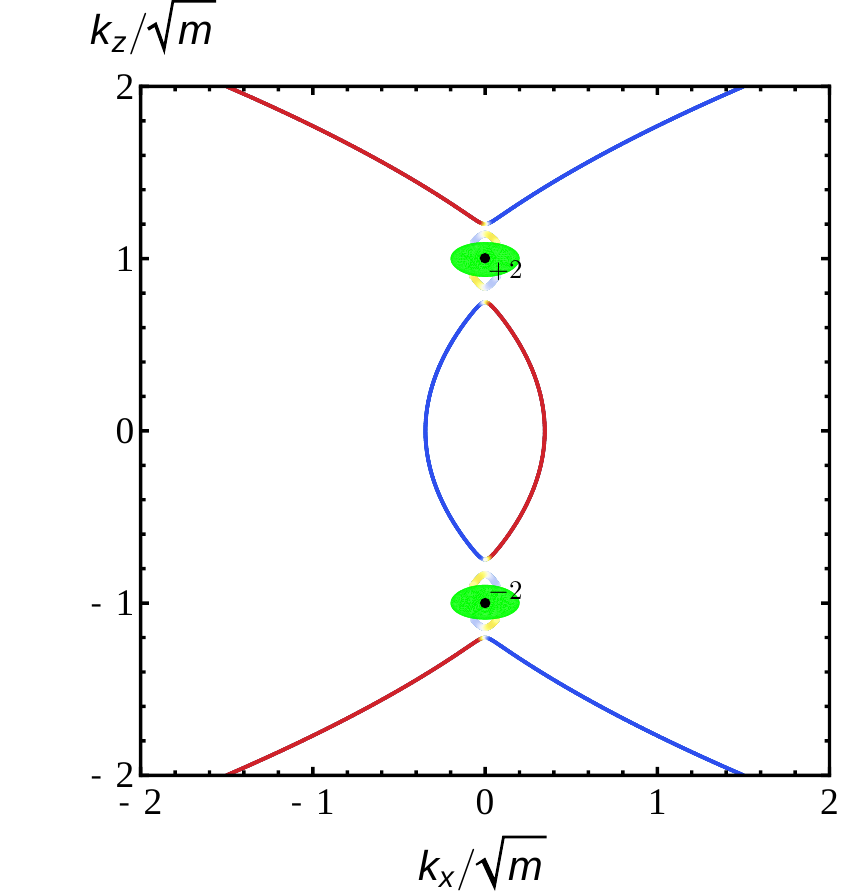}}
    \subfloat[$E=0.2 \, \gamma m$, $L=5/\sqrt{m}$ \label{vanished}] {\includegraphics[width=0.3\textwidth]{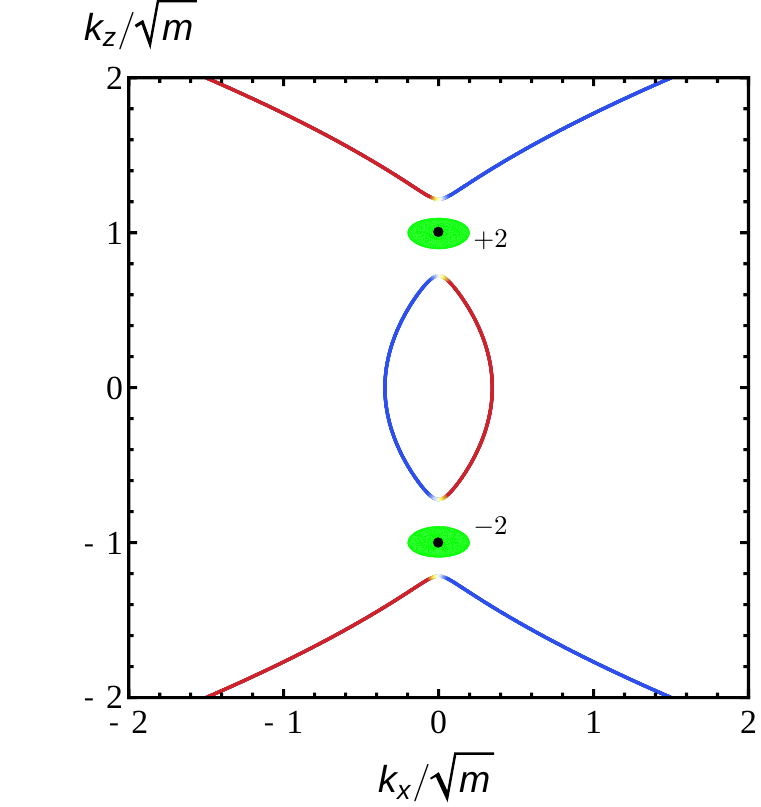}}
    \subfloat[$E=0.2 \, \gamma m$, $L=0.7/\sqrt{m}$ \label{centralvanished}] {\includegraphics[width=0.345\textwidth]{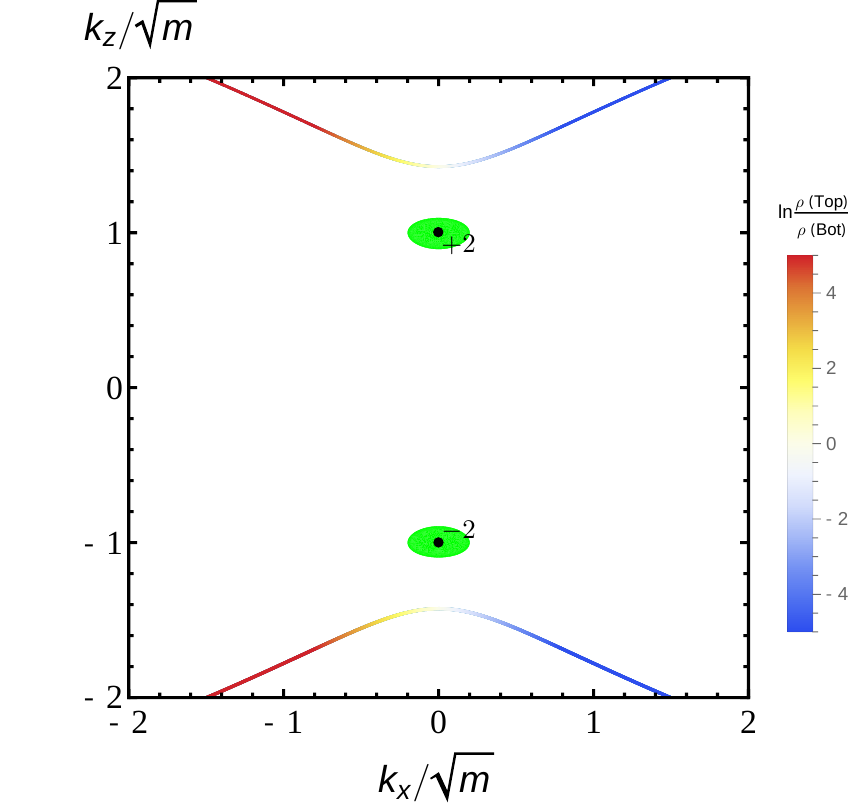}}
    \\
    \subfloat[$E=0.6 \, \gamma m$, $L=3/\sqrt{m}$ \label{fig:Thinarcs-noloop}] {\includegraphics[width=0.3\textwidth]{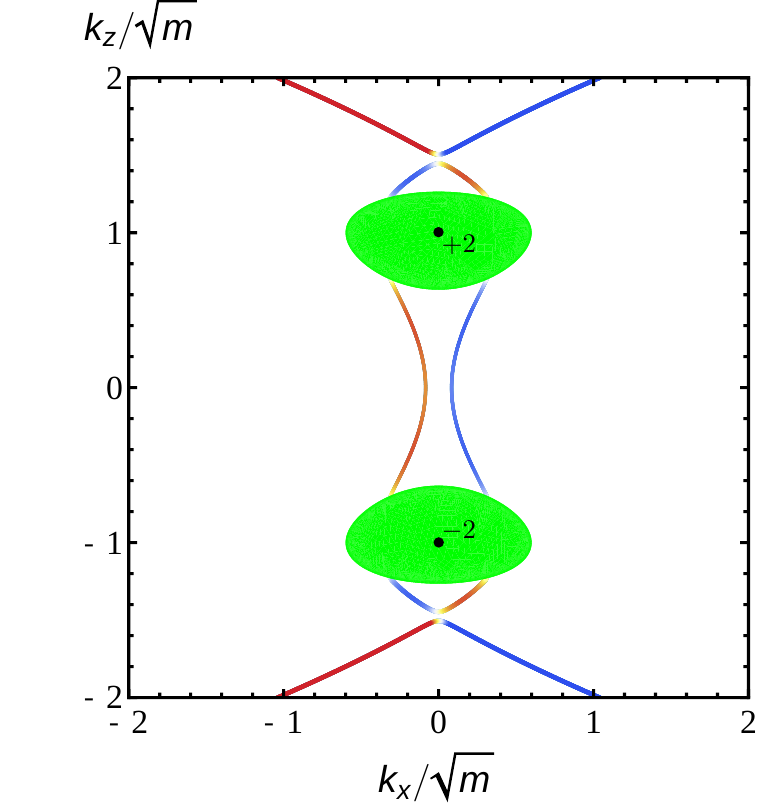}}
    \subfloat[$E=0.6 \, \gamma m$, $L=2.5/\sqrt{m}$ \label{fig:Thinarcs-reconnect}] {\includegraphics[width=0.3\textwidth]{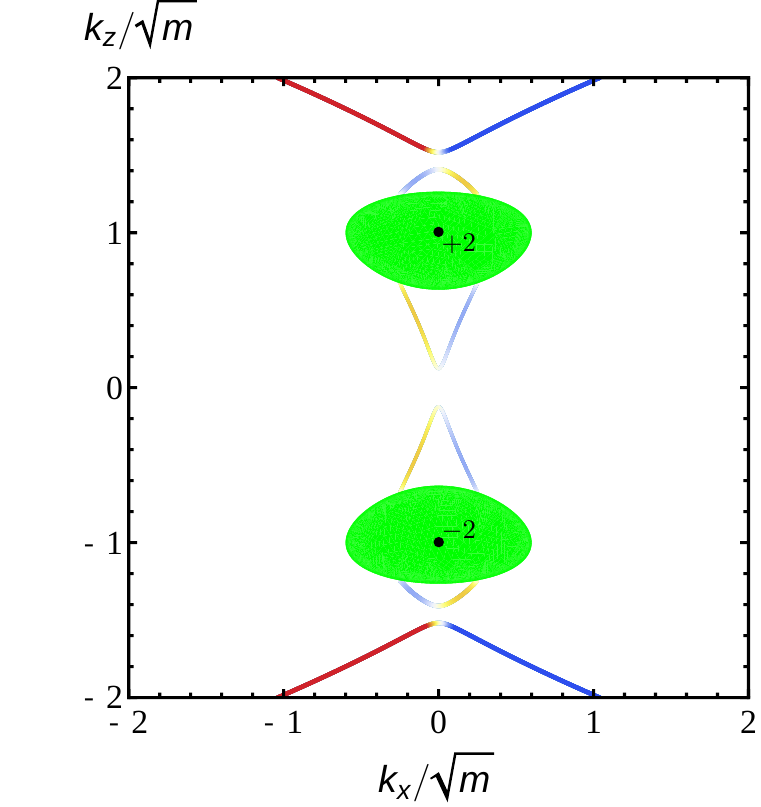}}
    \subfloat[$E=0.6 \, \gamma m$, $L=1.9/\sqrt{m}$ \label{fig:Thinarcs-nocentral}] {\includegraphics[width=0.345\textwidth]{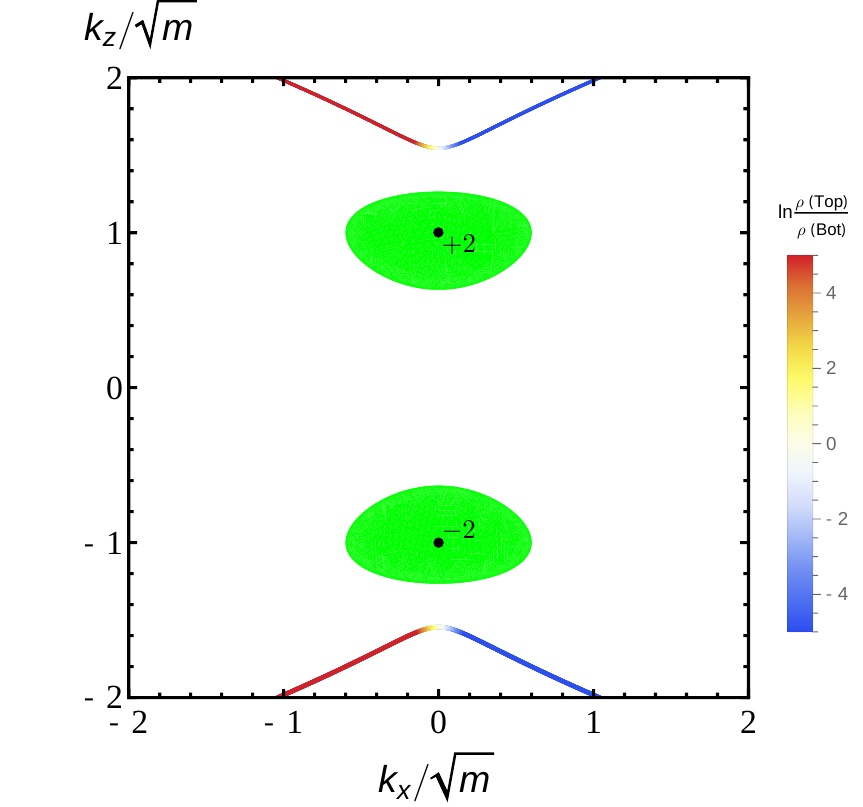}}
    \caption{
    Fermi arcs for a thin film of a triple-fold semimetal are shown as colored curves. The color indicates the spatial localization of the surface states: $\rho_t$ and $\rho_b$ are probability densities at the top and bottom surfaces, respectively. Points, whose value of $\ln\left(\rho_t/\rho_b\right)$ is above $4.5$ (below $-4.5$), are still marked as red (blue). Black dots and numbers $\pm2$ mark triple-fold crossing points and the corresponding topological charges.
    The projections of bulk states of semimetal at $L\to \infty$ are shown by green regions. In all panels, we fix $v_F=\gamma \sqrt{m}$.
    }
\label{fig:Thinarcs}
\end{figure*}

In the limit of a thick slab $L\rightarrow\infty$, we have $\coth(qL)\rightarrow1$ that gives
\begin{equation}
    1=\frac{2E^2-3v_F^2k_x^2+\gamma^2(k_z^2-m)^2}{2\sqrt{3}Ev_Fq},
\end{equation}
whose solutions reproduce the Fermi arcs described by Eq.~\eqref{true} and presented in Fig.~\ref{orc}.

Using Eq.~(\ref{Farcs}), we analyze how the thickness of the slab $L$ affects surface states. We plot the corresponding solutions for Fermi arcs in Fig.~\ref{fig:Thinarcs} for a few values of energy and slab thickness. For illustrative purposes, we also add the projections of the bulk states of the semimetal at $L\to \infty$ shown by green regions. The finite size effects lead to the separation of the Fermi arcs into several parts. The qualitative behavior and reconnection of the arcs depend on the energy $E$. We separate two parameter regions: (i) $0<E<\gamma m/2$ and (ii) $\gamma m/2 <E<\gamma m$. In the former case, the Fermi arcs at different surfaces of the thick slab have two intersections in momentum space at $k_z^2<m$, see Fig.~\ref{arc-E=0.2}. On the other hand, the arcs do not overlap in momentum space at $k_z^2<m$ for $\gamma m/2 <E<\gamma m$, see Fig.~\ref{arc-E=0.6}.

In the case $0<E<\gamma m/2$, the finite size effects lead to the hybridization of the Fermi arcs and their separation into several parts: two V-shaped parts at $k_z^2 >m$, petal-like parts that surround the bulk-state projections, and a closed loop at $k_z^2 <m$. For $L<L_{\text{cr,1}}$ with
\begin{equation}
    L_{\text{cr,1}}=\frac{2v_F}{\sqrt3 E},
\end{equation}
the petal-like branches vanish first while the internal loop remains, see Figs.~\ref{detached} and \ref{vanished}. By further reducing $L$, we remove the central part, see Fig.~\ref{centralvanished}. The corresponding critical length is determined by
\begin{equation}
    L_{\text{cr,2}}=\frac{v_F}{\sqrt{\gamma ^2m^2-E^2}}\artanh{\left(\frac{2\sqrt3 E\sqrt{\gamma^2m^2-E^2}}{2E^2+\gamma^2m^2}\right)}.
\end{equation}
The order in which the petal-like curves and the central loop disappear depends on a further subdivision of the energy range: below the energy $E_{\text{sim}}$, defined by $L_{\text{cr,1}}(E_{\text{sim}}) = L_{\text{cr,2}}(E_{\text{sim}})$, the petal-like curves vanish first; above $E_{\text{sim}}$, the central loop disappears first. We numerically estimate $E_{\text{sim}} \approx 0.43\, \gamma m$.

In the second case with $\gamma m/2 <E<\gamma m$, $L_{\text{cr,2}}$ defines a different lengthscale. The arcs reconnect in the central part for $L<L_{\text{cr,2}}$ and form petal-like structures surrounding the projections of the bulk states, see Fig.~\ref{fig:Thinarcs-reconnect}. Decreasing $L$ below $L_{\text{cr,1}}$ removes these petal-like structures, see Fig.~\ref{fig:Thinarcs-nocentral}.

Therefore, thin films of triple-fold semimetals are expected to show intricate patterns of the surface states.

\section{Finite-size effects}
\label{sec:doubly}

\subsection{Model}
\label{sec:doubly-model}

In the previous sections, we focused on the surface states in triple-fold semimetals. Let us now turn our attention to the finite-size effects for bulk states. For this purpose, it would be sufficient to consider only the low-energy states in the vicinity of the crossing points. Therefore, we adopt the following $\vec{k}\cdot \vec{p}$ Hamiltonian for pseudospin-1 fermions:
\begin{equation}\label{Prime}
    \mathcal{H}_0= v_F\begin{pmatrix}
        0 & i k_x & -ik_y\\
        -ik_x & 0 & ik_z\\
        ik_y & -ik_z & 0\\
    \end{pmatrix}= v_F(\vec{S_1}\cdot \vec{k}),
\end{equation}
where
\begin{equation}
\label{S-def}
\resizebox{.85\hsize}{!}{$S_{1x}=\begin{pmatrix}
        0 & i & 0\\
        -i& 0 & 0\\
        0 & 0 & 0\\
    \end{pmatrix},\, S_{1y}=\begin{pmatrix}
        0 & 0 & -i\\
        0& 0 & 0\\
        i & 0 & 0\\
    \end{pmatrix},\, S_{1z}=\begin{pmatrix}
        0 & 0 & 0\\
        0& 0 & i\\
        0 & -i & 0\\
    \end{pmatrix}$}
\end{equation}
are (pseudo)spin-1 matrices. In what follows, we will work with dimensionless parameters $E\rightarrow E/E_*$, $k\rightarrow k/k_*$ where $E_*$ and $k_*$ are characteristic energy and momentum. The characteristic momentum scale $k_*$ is chosen to be of the order of the inverse lattice constant, reflecting the typical scale at which the continuum approximation breaks down. The characteristic energy scale $E_*$ is taken to be of the order of the energy at which deviations from the effective $\vec{k}\cdot\vec{p}$ Hamiltonian approximation become significant. For example, $E_* \sim \gamma m$ for the model given in Eq.~\eqref{Hamiltonian-two-nodes}. To simplify our notations, we take $v_F=E_*/k_*$.

To address finite-size effects, we consider a semimetal-vacuum interface. As in Sec.~\ref{sec:Fermi-arcs}, we model electron states in vacuum by introducing a gapped Hamiltonian. However, like for a single Weyl node, it is impossible to introduce an energy gap for the $3\times 3$ Hamiltonian \eqref{Prime}. Therefore, we have to modify the model and consider a doubly degenerate triple-fold node. Before doing this, however, we find it instructive to prove why one cannot open a gap in the model given in Eq.~\eqref{Prime}.

A standard procedure to open a gap is to add a momentum-independent mass term to the Hamiltonian. The most general momentum-independent mass term in the case under consideration is given by the following Hermitian $3\times 3$ matrix:
\begin{equation}\label{MASS}
    \mathcal{H}_{m}=\begin{pmatrix}
        \mu_1 & \alpha & \beta\\
        \alpha & \mu_2 & \gamma\\
        \beta & \gamma & \mu_3\\
    \end{pmatrix},
\end{equation}
whose elements are real constants. In principle, the off-diagonal elements $\alpha$, $\beta$, and $\gamma$ could contain imaginary contributions; however, they can be easily absorbed by redefining the momenta components $k_x$, $k_y$, and $k_z$ in the low-energy Hamiltonian (\ref{Prime}). Eigenvalues of $\mathcal{H}_0+\mathcal{H}_{m}$ are determined by the following characteristic equation:
\begin{widetext}
\begin{eqnarray}
&&-E^3+E^2(\mu_1+\mu_2+\mu_3)+E(k^2+\alpha^2+\beta^2+\gamma^2-\mu_1\mu_2-\mu_2\mu_3-\mu_1\mu_3)\nonumber\\
&&+(\mu_1\mu_2\mu_3-\mu_1\gamma^2-\mu_2\beta^2-\mu_3\alpha^2+2\alpha\beta\gamma)-\mu_1k_z^2-\mu_2k_y^2-\mu_3k_x^2-2\beta k_xk_z-2\alpha k_yk_z-2\gamma k_xk_y=0.
\label{eigenvalue-equation}
\end{eqnarray}
\end{widetext}

Since we can always shift the reference point for energy in Eq.~(\ref{MASS}), it is sufficient to show that a gap cannot be opened at $E=0$. Therefore, it is sufficient to consider Eq.~\eqref{eigenvalue-equation} at $E=0$,
\begin{eqnarray}\label{surf}
&&\mu_1k_z^2+\mu_2k_y^2+\mu_3k_x^2+2\beta k_xk_z+2\alpha k_yk_z+2\gamma k_xk_y \nonumber\\
&&=\mu_1\mu_2\mu_3-\mu_1\gamma^2-\mu_2\beta^2-\mu_3\alpha^2+2\alpha\beta\gamma.
\end{eqnarray}
The gap is absent if Eq.~\eqref{surf} has a solution for a real-valued $\mathbf{k}$. To show that this is indeed the case, we rewrite Eq.~(\ref{surf}) as follows:
\begin{equation}\label{again}
    k_i(\mathcal{H}_{m})_{ij}k_j=\text{det}\,\mathcal{H}_{m}.
\end{equation}
Since det$(\mathcal{H}_{m})=\lambda_1\lambda_2\lambda_3$ is given by the product of eigenvalues of $\mathcal{H}_{m}$, applying an orthogonal transformation $O$, which diagonalizes matrix $\mathcal{H}_{m}$ in Eq.~\eqref{again}, we arrive at
\begin{equation}
\lambda_1\tilde{k}_z^2+\lambda_2\tilde{k}_y^2+\lambda_3\tilde{k}_x^2=\lambda_1\lambda_2\lambda_3,
\end{equation}
where $\tilde{\mathbf{k}}=O\mathbf{k}$. It is easy to verify that there is always a solution to the above equation for any choice of $\lambda_1$, $\lambda_2$, and $\lambda_3$. Therefore, we conclude that the gap cannot be opened at a single node of triple-fold semimetal via the mass term \eqref{MASS}.

However, a gap can be opened by considering a system with degenerate bands and adding terms to the Hamiltonian that mix these bands. We consider a pseudospin-1 semimetal with doubly degenerate bands defined by the Hamiltonian
\begin{equation}\label{Doublydegenerate}
   \mathcal{H}_{1}=\begin{pmatrix}
       \mathcal{H}_{0} & 0\\
       0 & \mathcal{H}_{0}^*
   \end{pmatrix}= (\vec{S_1}\cdot \vec{k})\otimes\tau_z.
\end{equation}
To model the vacuum surrounding the semimetal, we introduced an off-diagonal mass term
\begin{equation}
\label{Degeneratemass}
\mathcal{H}_{1}^{\text{gap}}=\mathcal{H}_{1}+\mu I_3\otimes\tau_1.
\end{equation}
This model is particularly convenient since its wave functions have the same structure as in the semimetal, albeit the energy spectrum is gapped as in the vacuum.

As we discussed in Sec.~\ref{sec:Fermi-arcs-matching}, only four out of six components of wave functions should be matched at boundaries. For the surface of the semimetal at $y=0$, we derive the following dispersion relation for the surface states:
\begin{equation} \label{Surfsemiinfinite}
    4E^2(k_x^2+k_z^2-E^2)^2=\mu^2(k_x^2+k_z^2-2E^2)^2.
\end{equation}
In the $\mu\rightarrow\infty$ limit, we obtain
\begin{equation}
    E=\pm\frac{k_{\parallel}}{\sqrt{2}},
\end{equation}
where $k_{\parallel}=\sqrt{k_x^2+k_z^2}$. Therefore, unlike the case in Sec.~\ref{sec:Fermi-arcs}, we have a circular surface state.

\subsection{Characteristic equation and energy spectrum}
\label{sec:doubly-spectrum}

To address the finite-size effects, we consider a slab of semimetal embedded in vacuum whose electron states are described by Hamiltonians~(\ref{Doublydegenerate}) and (\ref{Degeneratemass}), respectively, with the same orientation of the slab as in Sec.~\ref{sec:finite-size1}. In this setup, the semimetal is at $-L/2 < y < L/2$, hence $k_x$ and $k_z$ are good quantum numbers. The inverse decay lengths are determined by $q=\sqrt{k_x^2+k_z^2-E^2}$ and $p=\sqrt{\mu^2-E^2+k_x^2+k_z^2}$.

By matching the wave functions at boundaries and separating decaying in the bulk states from the propagating ones, we obtain the following characteristic equations for the surface states:
\begin{equation}\label{sixfoldresult}
2\coth\left(qL\right)=\frac{E}{\epsilon}\frac{p^2E^2-q^2\epsilon^2}{qpE\epsilon}\pm\frac{\mu}{\epsilon}\frac{p^2E^2+q^2\epsilon^2}{qpE\epsilon},
\end{equation}
where $\epsilon=\sqrt{\mu^2-E^2}$. The characteristic equation for bulk states can be obtained by substituting $q=il$, $l=\sqrt{E^2-k_x^2-k_z^2}$ in Eq.~(\ref{sixfoldresult}), leading to
\begin{equation}\label{cotlK}
2\cot\left(lL\right)=\frac{E}{\epsilon}\frac{p^2E^2+l^2\epsilon^2}{lpE\epsilon}\pm\frac{\mu}{\epsilon}\frac{p^2E^2-l^2\epsilon^2}{lpE\epsilon}.
\end{equation}

In the limit $\mu\rightarrow\infty$, the characteristic equation is reduced to
\begin{eqnarray}
\label{reducedsixfsol-1}
\coth (L\sqrt{k_x^2+k_z^2-E^2}) &=& \pm \frac{k_x^2+k_z^2}{2E \sqrt{k_x^2+k_z^2-E^2}},\\
\label{reducedsixfsol-2}
\cot (L\sqrt{E^2-k_x^2-k_z^2}) &=& \pm \frac{k_x^2+k_z^2}{2E \sqrt{E^2-k_x^2-k_z^2}}
\end{eqnarray}
at $E^2<k_x^2+k_z^2$ and $E^2>k_x^2+k_z^2$, respectively.

The dispersion relation of the flat band is not modified in this model, and the only finite-size effect of the system is the quantization of momentum component perpendicular to the surfaces $k_y= \pi n/L$, $n\in\mathbb{N}$.

\begin{figure*}[t]
    \centering
    \subfloat[$L=10$] { \includegraphics[width=0.47\textwidth]{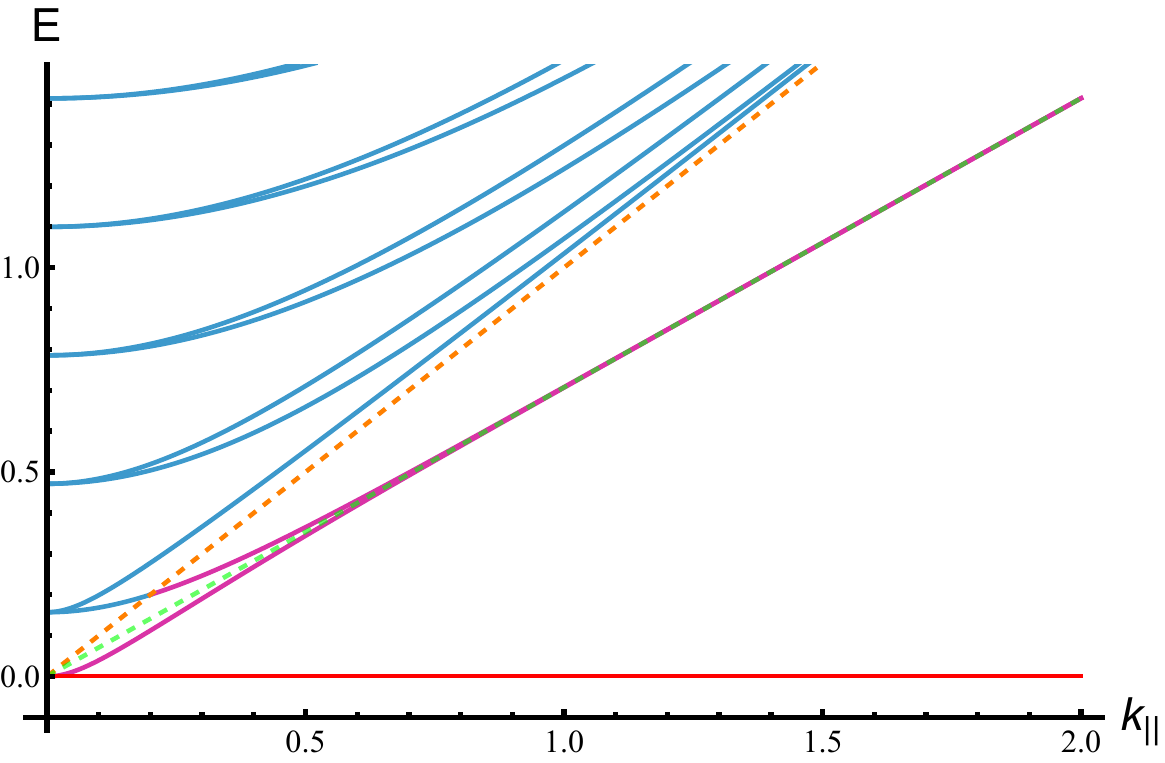}}
    \subfloat[$L=2$] { \includegraphics[width=0.47\textwidth]{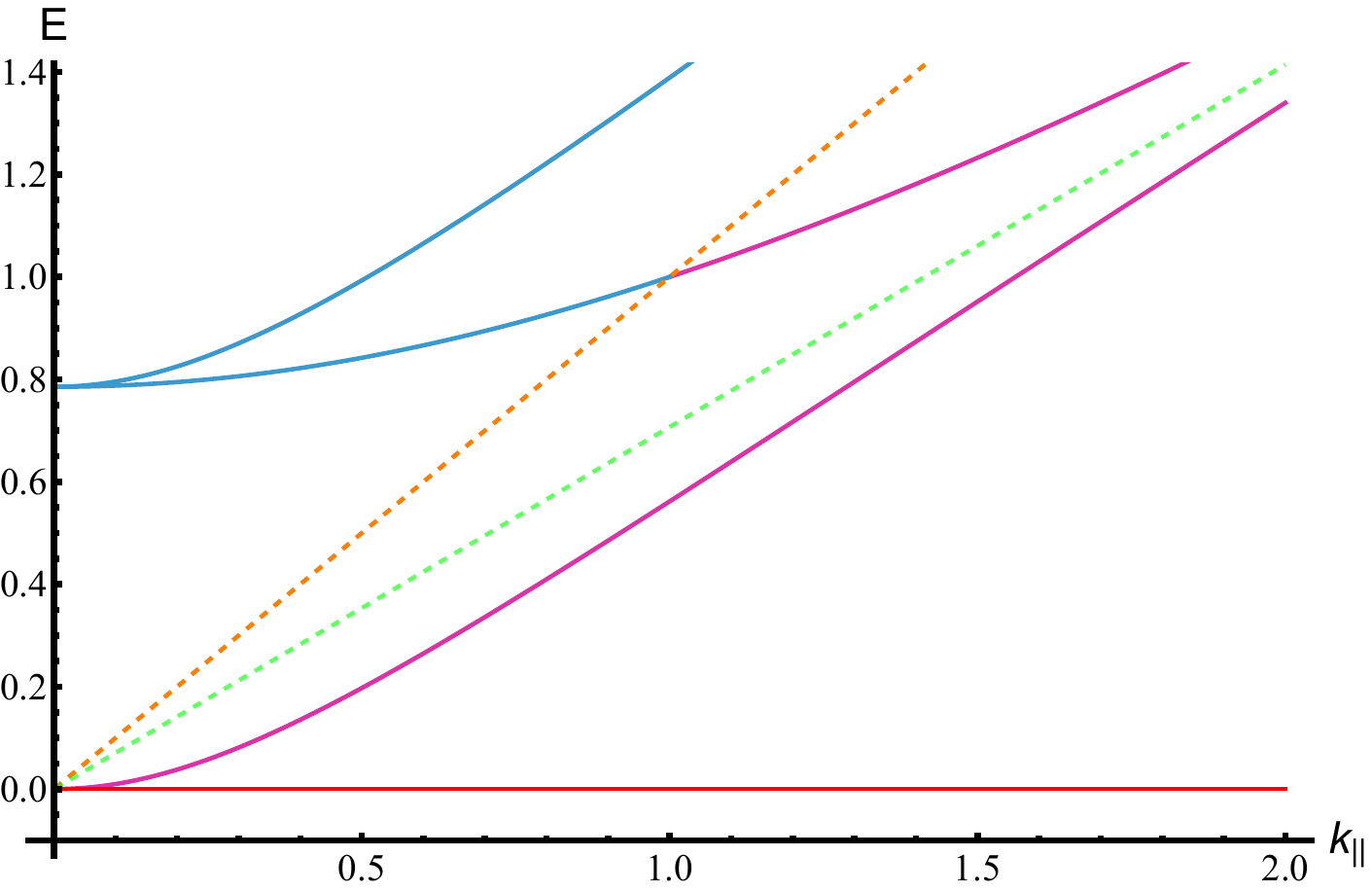}}
   ~
    \caption{
    Energy spectrum ($E\geq 0$) of a slab of doubly-degenerate triple-fold semimetal with finite thickness $L$ at $\mu \to \infty$ defined by Eqs.~\eqref{reducedsixfsol-1} and \eqref{reducedsixfsol-2}. The states above the orange dotted line correspond to bulk states (blue lines), while below that line, only surface states are possible, which are plotted in magenta.
    The flat band is plotted in red. The green dotted line represents the surface Dirac cone, which appears in the $L\rightarrow\infty$ limit.}
    \label{fig:L=10}
\end{figure*}

The energy spectrum defined by Eqs.~\eqref{reducedsixfsol-1} and \eqref{reducedsixfsol-2} is plotted in Fig.~\ref{fig:L=10} for $L=10$ (measured in the units of $1/k_{*}$). The bulk energy spectrum is quantized with the finite-size gap given by $\Delta E_{\text{gap}} = \pi/(2L)$. The bulk states appear in pairs of branches, corresponding to the $\pm$ sign in Eq.~(\ref{cotlK}). The distance between quantized energy levels is given by $\pi/L$. While most of the bulk energy levels extend indefinitely, the ground energy level deviates from this behavior: at $k_{\parallel} = 2/L$, it smoothly connects to the surface-state branch. We also note that there are always two branches of surface-localized modes. The key distinction between them lies in the internal structure of the wave function. To make this explicit, consider the six-component (spinor) functions $\psi_{\text{t}}\propto e^{qy}$, localized on the top surface, and, respectively, $\psi_{\text{b}}\propto e^{-qy}$, localized on the bottom surface. These functions are related by the symmetry transformation
\begin{equation}
    \psi_{\text{t}}=\begin{pmatrix}
    0 & I_3\\
    I_3 & 0
\end{pmatrix}\hat{T}\hat{E}\hat{M}_y\psi_{\text{b}},
\end{equation}
where $\hat{T}$ is the time reversal operator, $\hat{E}$ changes the sign of energy, $\hat{M}_y$ is the mirror reflection about the $y=0$ plane, and the prefactor matrix represents the valley-exchange operation. One can verify that this relation implies $\psi_{\text{b}}^\dagger\psi_{\text{b}}e^{2qy}=\psi_{\text{t}}^\dagger\psi_{\text{t}}e^{-2qy}$, preserving normalization under reflection.

The main distinction between the two branches, which correspond to the gapped (upper) and gapless (lower) surface states, lies in their construction. For the gapped branch, the total wave function within the region $-L/2 < y < L/2$ is given by $\Psi = \psi_{\text{b}} + \psi_{\text{t}}$, whereas for the gapless branch it takes the form $\Psi = \psi_{\text{b}} - \psi_{\text{t}}$. The explicit structure of $\psi_{\text{b}}$ and $\psi_{\text{t}}$ depends not only on the energy $E$ and magnitude of the in-plane momentum $k_{\parallel}$ but also on its direction.

The energy–momentum relation, expressed as a function of the magnitude of in-plane momentum $k_{\parallel}$, is characterized by families of curves that exhibit linear asymptotic behavior. Specifically, bulk states approach the asymptote $E=k_{\parallel}$, while surface-localized states approach $E = k_{\parallel}/\sqrt{2}$. Near $k_{\parallel}=0$, these curves can be well-approximated by a parabola. All bulk state branches eventually asymptotically approach the $E=k_{\parallel}$ line. Thus, branches with equal energy gaps (pairs of solutions of Eq.~(\ref{cotlK}) with $\pm$ sign) converge at $k_{\parallel}\rightarrow\infty$. Across the full momentum range, they are captured by branches of hyperbolas

\begin{widetext}
\begin{equation}\label{Hyperbolas}
E =\begin{cases}
    \frac12\sqrt{\frac{\pi^2}{L^2}+2k_{\parallel}^2}-\frac{\pi^2}{2L^2\sqrt{\frac{\pi^2}{L^2}+2k_{\parallel}^2}},&\text{for the gappless surface branch},\\
    \frac12\sqrt{\frac{\pi^2}{L^2}+2k_{\parallel}^2},&\text{for the ground-state bulk branch and the gapped surface branch},\\
    \frac12\sqrt{\frac{\pi^2(2n-1)^2}{L^2}+4k_{\parallel}^2},&\text{for the $n$-th pair of bulk branches}.
\end{cases}
\end{equation}
\end{widetext}

The hyperbolic approximation described above is applicable in the regions between each such pair of branches. This symmetric placement reduces error in estimating the density of states. The densities of the bulk and surface states are calculated in this approximation without taking the spin degeneracy into account, and the result reads

\begin{widetext}
 \begin{eqnarray}
 \label{DOS-bulk}
    N_{\text{bulk}}(E) &=& \frac{E}{\pi v_F^2}\left[\frac12\theta\left(E-\frac{\pi v_F}{2L}\right)-\theta\left(E-\frac{2 v_F}{L}\right)+  \sum\limits_{n=1}^\infty\theta\left(E-\frac{\pi(2n-1)v_F}{2L}\right)\right],\\
\label{DOS-surface}
    N_{\text{surface}}(E) &=& \frac{E}{\pi v_F^2}\theta\left(E-\frac{2 v_F}{L}\right)+\frac{1}{2\pi v_F^2}\Bigg(E+\sqrt{E^2+\frac{\pi^2 v_F^2}{L^2}}+\frac{E^2}{\sqrt{E^2+\frac{\pi^2 v_F^2}{L^2}}}\Bigg)\theta(E),
 \end{eqnarray}
\end{widetext}
where $\theta(x)$ denotes the unit step function and we restored $v_F$.

\section{Summary}
\label{sec:summary}

We analytically studied topological surface states in 3D triple-fold semimetals in a two-node continuum model, and addressed the finite-size effects in a continuum model of a doubly degenerate triple-fold semimetal.

It was found that the matching of wavefunctions at the semimetal-vacuum interface is peculiar for triple-fold semimetals in view of the presence of a flat band. In particular, we showed that only two out of three wavefunction components should be matched at the interface. Heuristically, this result follows from the vanishing normal component of the electric current because it involves only two out of three components of the wave functions. The same conclusion was reached by including a weak dispersion for the flat band.

In agreement with the topological charge arguments, we found that there exist two Fermi arcs per triple-fold node. While one Fermi arc connects two nodes, the other radiates away and is reminiscent of the long Fermi arc spanning the whole Brillouin zone observed in CoSi~\cite{Takane-Sato:2019, Rao-Ding:2019, Sanchez-Hasan:2018}, see Fig.~\ref{orc}.

In thin films of triple-fold semimetals, the surface states which belong to different surfaces hybridize, see Eq.~\eqref{film-char-eq} for the characteristic equation. Depending on energy and the semimetal thickness, a few scenarios are possible for the Fermi arc hybridization. For $0<E<\gamma m/2$, two petal-like curves connecting surface projections of bulk triple-fold nodes and a closed loop separated from the petal-like curves are realised, see Fig.~\ref{detached}. The central loop is absent for $\gamma m/2 <E<\gamma m$, see Fig.~\ref{fig:Thinarcs-noloop}.

Our study of finite-size effects in slabs of doubly degenerate triple-fold materials revealed that the dispersive bands become quantized with the distance between the levels determined by the thickness of the slab, see Eq.~\eqref{reducedsixfsol-2} for the corresponding characteristic equation. In addition to these bulk modes, we also found two surface-localized states with isotropic Dirac dispersion. One of these states is purely surface-localized, while the other transitions into the bulk state, see Fig.~\ref{fig:L=10}.

The obtained results will be instrumental for investigating thin films of multi-fold semimetals. The nontrivial shape of the arcs and the quantization of the bulk states can be accessed via angle-resolved photoemission spectroscopy (ARPES). Depending on the energy of photons, one can probe the dispersion relation of surface (low-energy photons) or bulk (high-energy photons, X-rays) states. Such an approach is widely utilized in topological materials where both surface and bulk states are observed~\cite{Hasan-Huang:rev-2017, Hasan-Yin:2021-rev, Sanchez-Barriga-Rader-AngleresolvedPhotoemissionTopological-2024}. Another experimental method, scanning tunneling spectroscopy~\cite{Fischer-Renner:review-2007}, can be used to extract information about the surface local density of states. While not being the focus of the current work, optical response could also be used to investigate the band structure of the material, see, e.g., Refs.~\cite{Flicker-Grushin:2018, Sanchez-Martinez-Grushin:2019, Ni-Wu-GiantTopologicalLongitudinal-2021} for the corresponding studies in multi-fold semimetals.

\begin{acknowledgments}
A.Yu.P. and E.V.G. acknowledge support from the National Research Foundation of Ukraine grant (2023.03/0097) “Electronic and transport properties of Dirac materials and Josephson junctions”.
P.O.S. thanks the Center for Quantum Spintronics (QuSpin) for warm hospitality, where this work was finalized.
\end{acknowledgments}

\appendix
\newpage

\begin{widetext}

\section{Berry curvature and Chern numbers}
\label{sec:app-topology}

To quantify the topological properties of the bulk model given in Eq.~\eqref{Hamiltonian-two-nodes}, we evaluate the Berry curvature at the band $n$ defined as
\begin{equation}\label{Curv}
\vec{\Omega}_n = -\text{Im}\sum_{i\neq n}\frac{\bra{\psi_n}(\vec{\nabla}_k H)\ket{\psi_i}\times\bra{\psi_i}(\vec{\nabla}_kH)\ket{\psi_n}}{(E_n-E_i)^2}
\end{equation}
and calculate its flux.

The eigenstates of Hamiltonian \eqref{Hamiltonian-two-nodes} read
\begin{equation}\label{newbasis}
    \psi_0=\frac{1}{E_+}\begin{pmatrix}
        \gamma(k_z^2-m)\\
        v_Fk_y\\
        v_Fk_x\\
    \end{pmatrix},\, \psi_+^{}=\sqrt{\frac{1}{2E_+^2(E_+^2-v_F^2k_x^2)}}\begin{pmatrix}
        v_Fk_x\gamma(k_z^2-m)+iv_Fk_yE_+\\
        v_F^2k_yk_x-i\gamma(k_z^2-m)E_+\\
        -E_+^2+v_F^2k_x^2\\
    \end{pmatrix},\, \psi_-^{}=\psi_+^{*},
\end{equation}
where $E_+=\sqrt{v_F^2(k_x^2+k_y^2)+\gamma^2(k_z^2-m)^2}$. The subscript corresponds to the flat band ($0$) and two dispersive bands ($\pm$).

Now we proceed with calculating the Berry curvature defined in Eq.~\eqref{Curv}. The vector operator $\vec{\nabla}_k H$ is
\begin{equation}
    \vec{\nabla}_k H = \left\{v_FS_{1x}, v_FS_{1y}, 2\gamma k_zS_{1z} \right\},
\end{equation}
where the matrices $S_{1i}$ are defined in Eq.~\eqref{S-def}.

The matrix elements of these matrices for dispersive bands are
\begin{eqnarray}
\bra{\psi_+}S_{1x}\ket{\psi_-} &=& \frac{i}{2E_+^2(E_+^2-v_F^2k_x^2)}\Big\{\left[ v_Fk_x\gamma(k_z^2-m)-iv_Fk_yE_+\right]\left[v_F^2k_yk_x+i\gamma(k_z^2-m)E_+\right] \nonumber\\
&-& \left[v_F^2k_yk_x+i\gamma(k_z^2-m)E_+\right]\left[ v_Fk_x\gamma(k_z^2-m)-iv_Fk_yE_+\right]\Big\}=0,\\
\bra{\psi_+}S_{1y}\ket{\psi_-} &=& \bra{\psi_+}S_{1z}\ket{\psi_-}=0.
\end{eqnarray}

The only nonzero matrix elements are the ones including a dispersive band and the flat one,
\begin{eqnarray}
\bra{\psi_+}v_FS_{1x}\ket{\psi_0} &=& \frac{i}{E_+^2\sqrt{2(E_+^2-v_F^2k_x^2)}}\Big\{\left[ v_Fk_x\gamma(k_z^2-m)-iv_Fk_yE_+\right]v_Fk_y-\left[v_F^2k_yk_x+i\gamma(k_z^2-m)E_+\right]\gamma(k_z^2-m)\Big\}\nonumber\\
&=& \frac{i}{E_+^2\sqrt{2(E_+^2-v_F^2k_x^2)}}\left[-iv_F^2k_y^2E_+-i\gamma^2(k_z^2-m)^2E_+\right] =v_F\frac{\sqrt{E_+^2-v_F^2k_x^2}}{\sqrt{2E_+^2}}, \\
\bra{\psi_0}v_FS_{1x}\ket{\psi_+} &=& -v_F\frac{\sqrt{E_+^2-v_F^2k_x^2}}{\sqrt{2E_+^2}},
\end{eqnarray}

\begin{equation}
    \bra{\psi_+}v_FS_{1y}\ket{\psi_0}=v_F\frac{-v_F^2k_yk_x-i\gamma(k_z^2-m)E_+}{\sqrt{2E_+^2(E_+^2-v_F^2k_x^2)}},\qquad\qquad \bra{\psi_0}v_FS_{1y}\ket{\psi_+}=v_F\frac{v_F^2k_yk_x-i\gamma(k_z^2-m)E_+}{\sqrt{2E_+^2(E_+^2-v_F^2k_x^2)}},
\end{equation}

\begin{multline}
    \bra{\psi_+}2\gamma k_zS_{1z}\ket{\psi_0}=2\gamma k_z\frac{iv_Fk_yE_+-\gamma(k_z^2-m)v_Fk_x}{\sqrt{2E_+^2(E_+^2-v_Fk_x^2)}},\qquad\bra{\psi_0}2\gamma k_zS_{1z}\ket{\psi_+}=2\gamma k_z\frac{iv_Fk_yE_++\gamma(k_z^2-m)v_Fk_x}{\sqrt{2E_+^2(E_+^2-v_F^2k_x^2)}},
\end{multline}
and
\begin{equation}
    \bra{\psi_-}\vec{S}_{1}\ket{\psi_0}=\bra{\psi_0}\vec{S}_{1}\ket{\psi_+}.
\end{equation}

Therefore, we obtain
\begin{equation}
    \bra{\psi_+}\vec{S}_{1}\ket{\psi_0}\times \bra{\psi_0}\vec{S}_{1}\ket{\psi_+}=-\frac {iv_F^2} {E_+} \left\{2\gamma k_zk_x, 2\gamma k_zk_y, \gamma(k_z^2-m)\right\}
    \quad \to \quad
    \vec{\Omega}_{\pm} = \pm \frac {v_F^2}{E_+^3}  \left\{2\gamma k_z k_x, 2\gamma k_z k_y, \gamma(k_z^2-m) \right\}.
\end{equation}

To prove that the Fermi arcs may extend beyond the multi-fold nodes, let us now consider the surfaces with fixed $k_z$ and calculate the flux of Berry curvature through them. We have
\begin{eqnarray}
&&\int_{k_z=const} \vec{\Omega}_+\frac{d\vec{S}}{2\pi} =\int_0^{2\pi}\int_0^\infty \frac{v_F^2\gamma(k_z^2-m)}{\left[\gamma^2(k_z^2-m)^2+v_F^2\rho^2\right]^{3/2}}\rho \frac{d\phi d\rho}{2\pi} =\int_{\gamma^2(k_z^2-m)^2}^\infty\frac{\gamma(k_z^2-m)}{2\left[\gamma^2(k_z^2-m)^2+\rho^2\right]^{3/2}}d[\gamma^2(k_z^2-m)^2+v_F^2\rho^2] \nonumber \\
&&=-\frac{\gamma(k_z^2-m)}{\left[\gamma^2(k_z^2-m)^2+v_F^2\rho^2\right]^{1/2}}\bigg|_{0}^\infty = \sign{m-k_z^2}.
\end{eqnarray}
The Chern number of the surface with constant $k_z$ is $\pm1$ with the sign determined by whether $|k_z|$ is smaller or larger than $\sqrt{m}$. Hence, due to bulk-boundary correspondence, we expect a single Fermi arc at $|k_z|<\sqrt{m}$ and a single Fermi arc at $|k_z|>\sqrt{m}$.

To calculate the topological charge of the multi-fold nodes, we consider the vicinity of the nodes $\vec{k}= \left\{0,0,\pm \sqrt{m}\right\}$. Then, the Berry curvature reads
\begin{equation}
\vec{\Omega}_+\xrightarrow{k_z\rightarrow\sqrt{m}}\frac{2\gamma \sqrt{m}v_F^2}{(4\gamma^2 mk_z^2+v_F^2k_x^2+v_F^2k_y^2)^{3/2}} \vec{k}.
\end{equation}

To determine the topological charge of the multi-fold nodes, we integrate the Berry curvature over a small sphere surrounding the nodal point ($k=const\ll\sqrt{m}$). We obtain
\begin{eqnarray}
&&\int_{k=const} \vec{\Omega}_+\frac{d\vec{S}}{2\pi} =\int_0^{2\pi}\int_0^{\pi} \frac{2\gamma \sqrt{m}v_F^2\,(\vec{k}\cdot\hat{k}) k^2\sin(\theta)}{\left[(4\gamma^2m-v_F^2)k^2\cos^2(\theta)+v_F^2k^2\right]^{3/2}} \frac{d\phi d\theta}{2\pi}\nonumber\\
&&= -\int_{1}^{-1}\frac{2\gamma \sqrt{m}/v_F}{\left[(4\gamma^2m/v_F^2-1)\cos^2(\theta)+1\right]^{3/2}}d\cos(\theta) = \int_{-1}^1\frac{2\gamma \sqrt{m}/v_F}{\left[(4\gamma^2m/v_F^2-1)+1/X^2\right]^{3/2}}\frac{dX}{X^3}\nonumber\\
&&= \int_{1}^\infty\frac{2\gamma \sqrt{m}/v_F}{\left[(4\gamma^2m/v_F^2-1)+Y\right]^{3/2}}dY  = -2\frac{2\gamma \sqrt{m}/v_F}{\sqrt{(4\gamma^2m/v_F^2-1)+Y}}\Bigg|_{Y=1}^\infty=2.
\end{eqnarray}
For the node at $k_z = -\sqrt{m}$, the topological charge has the opposite sign.
Therefore, the topological charge of the triple-fold nodes is $\pm 2$.

\section{Wave function matching}
\label{sec:app-A}

In this appendix, we provide technical details of the wave function matching at the interface between the triple-fold semimetal and the vacuum. We use the model Hamiltonians given in Eqs.~(\ref{Hamiltonian-two-nodes}) and (\ref{Twonodevac}) in the triple-fold semimetal and the vacuum, respectively.

The eigenstates of the system with the energy $E$ are defined by
\begin{equation}\label{sysag}
    \begin{pmatrix}
        \gamma\tilde{M}(y) & i v_F k_x & -v_F \partial_y\\
        -i v_F k_x & \gamma\tilde{M}(y) & i\gamma\left[k_z^2-\tilde{m}(y)\right]\\
        v_F \partial_y & -i\gamma\left[k_z^2-\tilde{m}(y)\right] & \gamma\tilde{M}(y)\\
    \end{pmatrix}\Psi=E\Psi
\end{equation}
with the position-dependent parameters
\begin{equation}\label{app-my}
\tilde{m}(y)=\begin{cases}
    -M, & y>0,\\
    m, & y<0,
\end{cases}\qquad\qquad \tilde{M}(y)=\begin{cases}
    \frac M2, & y>0,\\
    0, & y<0.
\end{cases}
\end{equation}
In view of the translation invariance in the $x$ and $z$ directions, we seek eigenstates as plane waves with the momenta $k_x$ and $k_z$. Then the general solution to Eq.~(\ref{sysag}) for bulk states is given by
\begin{equation}\label{app-psi}
\Psi=e^{i(k_xx+ik_zz)}\begin{cases}
    C_1\begin{pmatrix}
        v_F k_x\gamma(k_z^2-m)+iv_F qE\\
        v_F^2 qk_x-i\gamma(k_z^2-m)E\\
        -E^2+v_F^2 k_x^2\\
    \end{pmatrix}e^{iqy}+C_2\begin{pmatrix}
        v_F k_x\gamma(k_z^2-m)-i v_F qE\\
        -v_F^2 qk_x-i\gamma(k_z^2-m)E\\
        -E^2+v_F^2 k_x^2\\
    \end{pmatrix}e^{-iqy}, & y<0,\\
    C_3\begin{pmatrix}
        v_F k_x\gamma(k_z^2+M)- v_F p \left(E-\gamma\frac M2\right)\\
        iv_F^2 pk_x-i\gamma(k_z^2+M) \left(E-\gamma\frac M2\right)\\
        -\left(E-\gamma\frac M2\right)^2+v_F^2 k_x^2\\
    \end{pmatrix}e^{-py}, & y>0,
\end{cases}
\end{equation}
where $C_1$, $C_2$, $C_3$ are arbitrary constants and
\begin{equation}\label{app-qp-def}
v_F q=\sqrt{E^2-\gamma^2\left(k_z^2-m\right)^2-v_F^2 k_x^2}, \qquad v_F p=\sqrt{\gamma^2\left(k_z^2+ M\right)^2+v_F^2k_x^2-\left(E-\gamma\frac{M}{2}\right)^2}.
\end{equation}
Matching all three components of the wave function at the interface $y=0$ results in the following system of three linear homogeneous equations for $C_1$, $C_2$, and $C_3$:
\begin{equation}\label{linsystem}
\begin{aligned}
        &\big(-E^2+v_F^2k_x^2\big)(C_1+C_2)=\left[ -\left(E-\gamma\frac M2\right)^2+ v_F^2k_x^2\right]  C_3,\\
        &k_x\gamma(k_z^2-m)(C_1+C_2)+iqE(C_1-C_2)=\left[k_x\gamma(k_z^2+M)-p\left(E-\gamma\frac M2\right) \right]C_3,\\
        &v_F^2qk_x(C_1-C_2) -i\gamma(k_z^2-m)E(C_1+C_2)=\left[iv_F^2 pk_x-i\gamma(k_z^2+M)\left(E-\gamma\frac M2\right)\right]C_3.
\end{aligned}
\end{equation}
Nontrivial solutions to this system exist when its determinant equals zero. In the limit $M \to \infty$, the determinant takes the form
\begin{eqnarray}\label{Delta}
    &&q\Bigg\{v_F^2 k_x^2\gamma(k_z^2+M) + v_F^2 pk_x\gamma\frac M2-E\gamma(k_z^2+M)\left(E-\gamma\frac M2\right)
    -\gamma(k_z^2-m)\left[v_F^2 k_x^2 -\left(E-\gamma\frac M2\right)^2\right] \Bigg\}
    \nonumber\\
&&\xrightarrow{M\rightarrow\infty}\frac14\gamma^2M^2 q \left[ 2E+\sqrt{3}v_F k_x+\gamma(k_z^2-m) \right].
\end{eqnarray}
This determinant vanishes for $q=0$, leading to a solution that grows linearly with $y$ and is not normalized. As a result, the solution for $y<0$ contains only one arbitrary constant, making it trivial to show that the matching conditions cannot be satisfied. At $q\neq0$, the expression in the square brackets vanishes only at $E=-\left[\sqrt{3} v_F k_x+\gamma(k_z^2-m)\right]/2$. However, in this case,
\begin{equation*}
    v_F q=\sqrt{\frac34 v_F^2 k_x^2+\frac{\sqrt{3}}{2} v_F k_x\gamma(k_z^2-m)+\frac14\gamma^2(k_z^2-m)^2-\gamma^2\left(k_z^2-m\right)^2-v_F^2 k_x^2}=\frac12\sqrt{-\left[\sqrt{3}\gamma\left(k_z^2-m\right)- v_Fk_x\right]^2}
\end{equation*}
is purely imaginary; hence, it cannot correspond to bulk states. Furthermore, in view of the structure of the wave function in Eq.~\eqref{app-psi}, this solution does not correspond to normalizable surface states.

Thus, the system admits only trivial solutions, which shows that the matching of all components of the wave function across the semimetal-vacuum interface is not possible.

\section{Dependence of surface spectrum on mass parameter}
\label{sec:app-B}

Let us address the dependence of the dispersion relation of the thin-film surface states given in Eq.~(\ref{Surfsemiinfinite}) on the mass parameter $\mu$. This parameter should be of the order of the work function, which is typically in the eV range. While typical values of characteristic energy should be in the range of $1$ meV - $100$ meV, it is important to verify that the limit $\mu\rightarrow\infty$ yields reliable results within this interval.

We compare the energy spectrum of the surface states in the infinite gap limit $\mu\rightarrow\infty$ with its counterpart for a finite gap in Fig.~\ref{fig:MassEffect}. For realistic values of the gap, the surface state spectrum is in good agreement with its counterpart in the infinite mass limit.

\begin{figure}[h]
    \centering
    \subfloat[$\mu = 10 E_*$]{\includegraphics[width=0.45\textwidth]{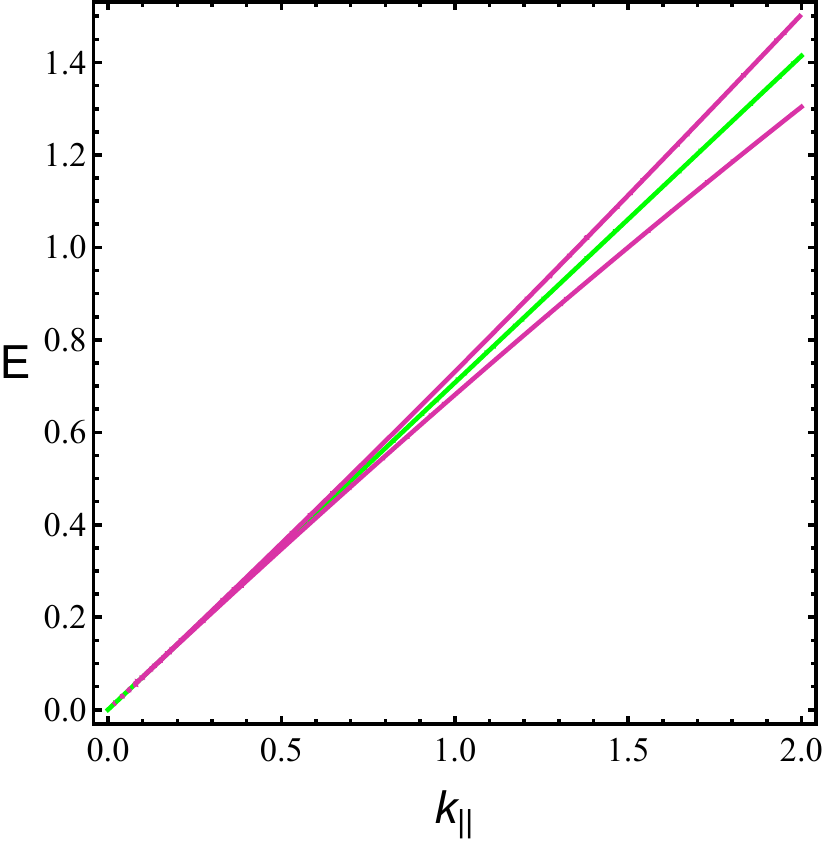}}
    \subfloat[$\mu = 30 E_*$]{\includegraphics[width=0.45\textwidth]{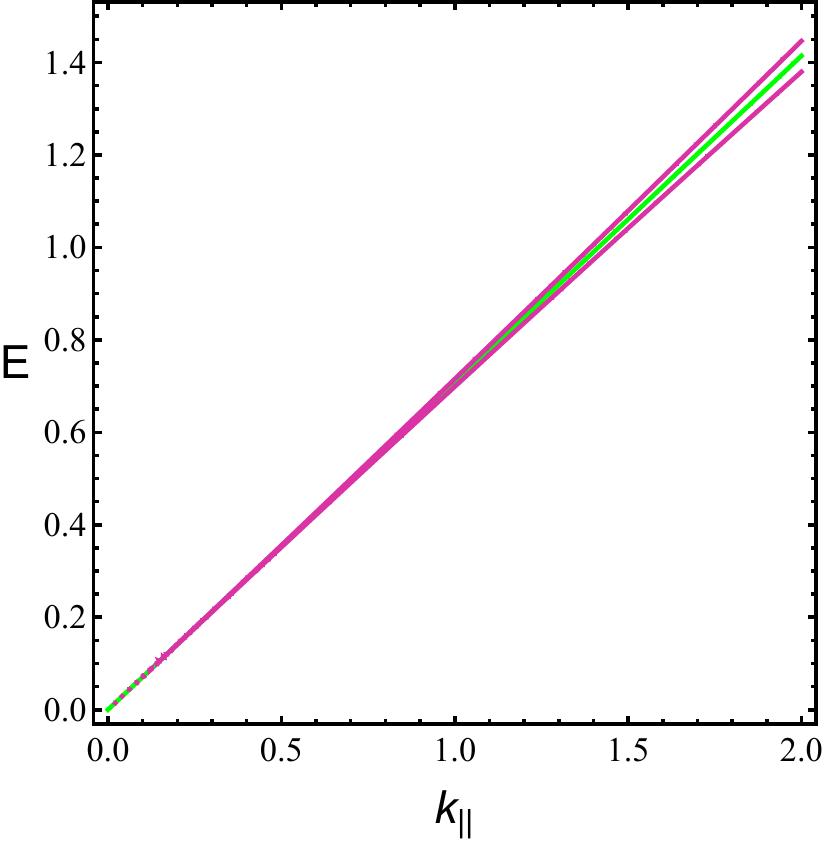}}~
    \caption{
    Dispersion relation of surface states given in Eq.~\eqref{Surfsemiinfinite} for finite values of gap parameter $\mu$ (pink) and at $\mu\rightarrow\infty$ (green). We use the dimensionless variables, see the text after Eq.~\eqref{S-def} for the definition.
    }
    \label{fig:MassEffect}
\end{figure}

As is evident from Fig.~\ref{fig:MassEffect}, for small values of momentum, where the finite-size effects are the strongest, the influence of the finite gap is negligible. Therefore, the infinite mass assumptions made during the investigation of surface states and finite size effects were justified.

\end{widetext}

\bibliography{Library-short}

\end{document}